\documentstyle[aps,epsfig,twocolumn]{revtex}

\begin{document}
\draft
\title{
Simulational Study on Dimensionality-Dependence of Heat Conduction
}
\author{Takashi Shimada\footnote{shimada@acolyte.t.u-tokyo.ac.jp}, 
Teruyoshi Murakami\footnote{murakami@acolyte.t.u-tokyo.ac.jp}, 
Satoshi Yukawa\footnote{yukawa@ap.t.u-tokyo.ac.jp},
Keiji Saito\footnote{saitoh@spin.t.u-tokyo.ac.jp},
and 
Nobuyasu Ito\footnote{ito@catalyst.t.u-tokyo.ac.jp}}
\address{Department of Applied Physics, 
School of Engineering, \\ The University of Tokyo, 
Bunkyo-ku, Hongo, Tokyo 113-8656, Japan}
\date{\today}
\maketitle
\begin{abstract}
Heat conduction phenomena are studied theoretically using computer
simulation. The systems are crystal with nonlinear interaction, and 
fluid of hard-core particles. Quasi-one-dimensional system of the size
of $L_x\times L_y\times L_z(L_z\gg L_x,L_y)$ is simulated. Heat baths
are put in both end: one has higher temperature than the other. 
In the crystal case, the interaction
potential $V$ has fourth-order non-linear term in addition to the harmonic
term, and Nose-Hoover method is used for the heat baths. 
In the fluid case, stochastic boundary condition is charged, which
works as the heat baths. 
Fourier-type heat conduction is reproduced both in crystal and fluid
models in three-dimensional system, but it is not observed in lower
dimensional system. 
Autocorrelation function of heat flux is also observed and long-time
tails of the form of $\sim t^{-d/2}$, where $d$ denotes the
dimensionality of the system, are confirmed. 
\end{abstract}
\noindent
\pacs{PACS number:
05.70.Ln
05.45.-a
05.20.Jj
}

The normal heat conduction, which is described by the Fourier heat law
\begin{equation} 
J=- \kappa \cdot \mbox{grad} (T) 
\label{fourier}
\end{equation}
where $J$ and $T$ denote the heat flux and the temperature respectively,
is one of the most fundamental nonequilibrium phenomena.
However the microscopic origin of it has not been clarified completely.
Unlike the Debye's theory for the specific heat in equilibrium solids,
this phenomena cannot be modeled by the harmonic chain\cite{RIEDER}.
In the harmonic chain, global internal temperature gradient is not formed, so that the thermal conductivity diverges.
Disorder of mass of particles in the system does not give the saturated thermal conductivity in the thermodynamic limit\cite{OCONNOR}.
This behavior attributes to the lack of scattering process between modes in the system.
On the other hand, some chaotic systems realize the Fourier heat law\cite{DING-A-LING,DINGDONG,FK}.
Thus the {\it nonintegrability} of the system which causes scattering between modes is one of crucial elements for the Fourier heat law.
It is worth stressing that all these chaotic models and the Lorentz-gas model, 
which is used as the model of the heat conduction in metals or diluted gas\cite{Lorentzgas1,Lorentzgas2}
are characterized by the nature that the total momentum is not conserved in the bulk.

Recent studies for total momentum conserving system by Lepri {\it et al.}
shed a light on another crucial aspect, that is, {\it dimensionality}\cite{LEPRIA}.
They first study the thermodynamic behavior of energy transport in the one-dimensional Fermi-Pasta-Ulam (FPU) $\beta$ 
lattice numerically and analytically investigated the dimensionality-dependence.
In the numerical investigation for the one-dimensional case, they found
numerically that the thermal conductivity diverges as $ N^\alpha $ 
where $N$ is the number of oscillators and $ \alpha $ is estimated roughly to be $1/2$.
They also calculated the power spectrum $S( \omega )$ of the global heat flux in the equilibrium 
and found its power-law divergence as ${\omega}^{-0.37}$ in the low-frequency region.
It corresponds to a slow decay of autocorrelation function of globally averaged heat flux $C_j (t)$ as $t^{-0.63}$.
This power-low decay of the autocorrelation function implies the divergence of the thermal conductivity
as $ \kappa \sim N^{\alpha \approx 0.37} $ 
by applying the Green-Kubo formula\cite{Kubo1,Kubo2}
\begin{equation}
  \kappa = \frac{1}{k_{\rm B} T^2} \int_0 ^\infty C_j (t) dt ,
  \label{GKF}
\end{equation}
because the upper limit of the time integration is expected to be proportional to the system size.
Moreover they took an analogy with the hydrodynamical systems
where the velocity autocorrelation function decays slowly as $t^{-d/2}$ ($d$ is the dimensionality of the system) 
when the density is high\cite{Alder1,Alder2,DorfmanCohen1,DorfmanCohen2,Ernst}.
In those systems, the autocorrelation function of heat flux also has the long-time-tails of the same exponent.
From this analogy, 
they conjectured the logarithmic divergence in two-dimensional system and 
the converging thermal conductivity in three-dimensional system\cite{LEPRIB}.
Following these studies, 
Hatano reported a diverging result in the one-dimensional diatomic Toda lattice\cite{HATANO}
and recently 
Lippi and Livi confirmed their prediction for the two-dimensional
FPU-$\beta$ lattice and Lennard-Jones $6/12$ lattice\cite{LIPPI}.

Although the analogy with dense-fluids is plausible for one and two-dimensional system,
there has been no direct study on either three dimensional lattices or dense fluids system.
Thus in the following we investigate the dimensionality dependencies of heat conduction on 
three-dimensional nonlinear lattice and dense fluid of hard-core particle.

Our model for crystal is coupled three-dimensional FPU-$\beta$ lattice with Nos\'{e}-Hoover thermostats
at both sides.
Hamiltonian of this system is represented as follows,
\begin{eqnarray} 
{\cal H}_0 & = & \sum_{i=1}^N \frac{ \mbox{\boldmath $p$}_i^2}{2} + \sum_{i,j ; n.n} V_{ij}, \label{ham1} \\
V_{ij} & = & \frac{1}{2} ( dq_{ij} )^2 + \frac{g}{4} ( dq_{ij} )^4 \quad
 ( dq_{ij}  =  | \mbox{\boldmath $q$}_i - \mbox{\boldmath $q$}_j | -l_0 ),
\label{ham2}
\end{eqnarray}
where $N,l_0,g$ denote the number of particles, natural length of the bond 
and the coefficient of the fourth order nonlinear term respectively.
Mass of the particles are uniformly 1.
Here dynamical variables 
$\mbox{\boldmath $p$}_i$ $= (p_i^x,p_i^y,p_i^z)$ and $\mbox{\boldmath $q$}_i$ $= (q_i^x,q_i^y,q_i^z)$ are three-dimensional.
The summation over $ V_{ij} $ takes only for nearest neighbor pairs.
In this system, the total momentum is conserved in the bulk since 
potential is simply the function of distance between two particles.
We will study two kinds of lattices; one is one-dimensional-like $(1 \times 1 \times N_z)$ lattices, 
and the other is three-dimensional-like one $(3 \times 3 \times N_z)$ where $N_z$ is taken to be from $8$ to $256$.
In the following we will call the former as 1D-lattice and the latter as 3D-lattice; although the 3D-lattice is actually
quasi-one-dimensional, we will see this system maintains the three-dimensionality later.
We take periodic boundary condition to the bonds vertical to $N_Z$-direction.
Note that our model have no long-distance-correlations.
Nos\'{e}-Hoover thermostats\cite{NOSE,HOOVER} act on each particle in both ends of $N_Z$-direction to keep their temperature constant.
One end is kept to be higher temperature($T_L$) than the other($T_R$).
Therefore the equations of the motion of this system are 
\begin{equation} 
  \begin{array}{l}
    \mbox{\boldmath $\dot{q}$}_i = \mbox{\boldmath $p$}_i , \\
    \mbox{\boldmath $\dot{p}$}_i =
    \cases{
      - \frac{ \partial {\cal H}_0 }{ \partial \mbox{\boldmath $q$}_i} & \mbox{(in the bulk)} \cr \cr
      - \frac{ \partial {\cal H}_0 }{ \partial \mbox{\boldmath $q$}_i} - \zeta_i \mbox{\boldmath $p$}_i 
      & \mbox{(at the both ends)}}
    \label{ham3}
  \end{array}
\end{equation}
and thermostat variables move as follows;
\begin{equation} 
\dot{ {\zeta}_i } = \frac{1}{Q} ( \frac{p_i^2}{3k_{\rm B} T} - 1). 
\label{ham4}
\end{equation}
Here $T$ is the temperature of the heat bath($T_L$ or $T_R$) and
$Q$ is the parameter of coupling between the thermal bath and the system.
In this study we set $Q=1$ in order that the response time of the thermostats($\sim 1/\sqrt{Q}$)
becomes the same order as the original time scale of the lattice.
The value of other parameters are chosen as follows;
\begin{eqnarray*} 
k_{\rm B} = 1,\; T_L =152/3, T_R =8,\; g=0.1,\; l_0 =100.\nonumber
\end{eqnarray*}
Under this condition, system is fully chaotic\cite{Ford}.
The 3D-lattice corresponds to the van-der-Waals solids since 
the natural length is enough long comparing with the displacement of particles which is usually smaller than $10$,
in other words,
nonlinearity comes from the presence of the natural length is negligible,
however it changes the strongest coupling combination of the displacements. 
We used the fourth-order predictor-corrector method to integrate the differential equations.
Time step we use$(10^{-3})$ for integration provides the total energy conservation with a relative accuracy $10^{-7}$ 
in the isolated system for the time corresponds to the longest simulation.

We define the local temperature as
\begin{equation} 
T(i)=<\mbox{\boldmath $p$}_i^2/3>,
\label{tempdiff}
\end{equation}
where $< \cdot >$ means average over long time.
The local heat flux $j_{kl}$ is defined as the energy transfer per unit time 
from the $k$th particle to the nearest particle $j$ as 
\begin{equation} 
j_{kl} = - \mbox{\boldmath $p$}_k \cdot \frac{\partial V_{kl}}{\partial \mbox{\boldmath $q$}_l}.
\end{equation}
For the 3D-lattice, we will treat only fluxes flow along the $N_Z$-direction bonds.
Then we can represent the flux from the $k$th particle to the next particle in $N$-direction simply as $j_k$.
The average of these heat flux $j_{k}$ over time and whole system is calculated as
\begin{equation} 
<j> = <\frac{1}{N'} \sum_k^{N'} j_k>
\end{equation}
where $N'$ is equal to $(N-1)$ for 1D-lattice and $(N-9)$ for 3D-lattice.

It was confirmed that the temperature profile takes a linear form
and the effect of boundary with thermostats was negligible in both ends.
From this fact we reasonably define the global thermal conductivity as
\begin{equation} 
\kappa = \frac{<j>N_z}{T_L-T_R},
\end{equation}
by substituting $(T_L-T_R)/N_z$ for the temperature gradient in Eq.(\ref{fourier}).

\begin{figure}
\begin{center}
\rotatebox{-90}{\mbox{\epsfig{file=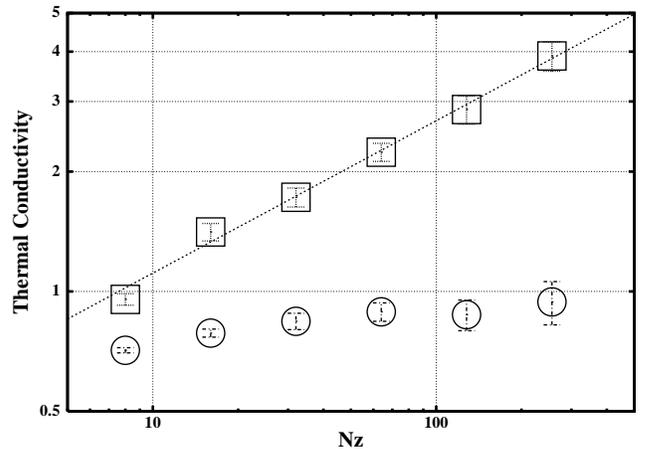,width=6cm}}}
\caption{Size dependence of the thermal conductivity in lattice systems is given in log scale.
Boxes denote the one in 1D-lattice $(1 \times 1 \times N_z)$ and circles denote the one in 3D-lattice $(3 \times 3 \times N_z)$.
Error bars show $2\sigma$ error regions.
The dotted line is proportional to $N_z^{0.38}$ which is obtained from a best fit 
by the power law for 5 points without the smallest system's one.
}
\label{kappaplot}
\end{center}
\end{figure}

Size-dependence of the thermal conductivity $\kappa$ is shown in FIG.\ref{kappaplot}.
In the 1D-lattice, thermal conductivity diverges as $ \sim N_z^{0.4} $ 
showing a good agreement with the study by Lepri {\it et al.}\cite{LEPRIA}.
However, in the 3D-lattice,
the result makes a sharp contrast with the 1D-lattice(FIG.\ref{kappaplot}).
The thermal conductivity converges
although the systems are actually quite narrow especially for large $ N_z $
and system become 1D-lattice like in the infinite size limit$(N_z \rightarrow \infty)$.

\begin{figure}
\begin{center}
\rotatebox{-90}{\mbox{\epsfig{file=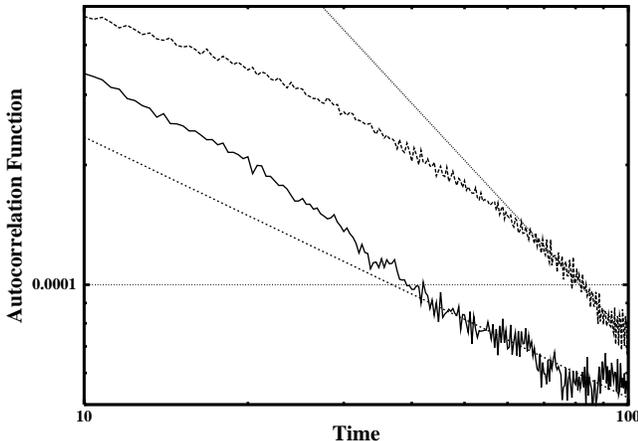,width=6cm}}}
\caption{The autocorrelation function of the isolated FPU lattices.
Solid line is the chain system $(1 \times 1 \times 64)$ and the dotted line is the one in the $(3 \times 3 \times 64)$ lattice.
Straight lines are obtained as the fitting in the long time region for power law.
The gradient of the fitting line for the 1D-system is about $ -0.65$
and the one for the 3D-system is about $-1.5$.}
\label{AC}
\end{center}
\end{figure}

In order to check the analogy with hydrodynamical system,
we also calculate temporal autocorrelation functions of average heat flux in the isolated systems.
In this case the boundary conditions are periodic for all directions.
These systems have no contact with the heat bath and the energy density
is chosen to the value corresponding to $(T_L + T_R)/2 = 88/3$.
In FIG.\ref{AC}, we can see that the autocorrelation function of 1D-system decays as $t^{-0.65}$ 
and the one of 3D-system decays $t^{-1.5}$ asymptotically.
Thus these results show the agreement with the Lepri {\it et al.}'s conjecture.

Next, to study the thermal conductivity of fluid, the system of hard-core particles is simulated.
The particles, with the uniform radius($\sigma=0.1$) and mass($m=1$), collide elastically with each other.
It is clear that this system also conserves the total momentum in the bulk.
The dimension of the objective systems are two$(L_y \times L_z)$, and three$(L_x\times L_y\times L_z)$
where $L_x$ and $L_y$ is fixed to 1 and $L_z$ is taken to be from 2 to 20.
The fixed width of the system($L_x,L_y$) is chosen from following condition;
it must be large enough comparing with the radius of the particles,
and on the other hand, it should be small to eliminate the hydrodynamic flow
since we are interested in the Fourier type heat conduction. 
The left and right sides of the system are the walls with temperature $T_L=6$ and $T_R=2$ (again $T_L > T_R$),
and the boundary conditions for other directions are periodic.
If a particle comes to the left or right end, where the stochastic heat bath is acting,
velocity is randomly chosen according to the equilibrium distribution
\begin{eqnarray}
f(\mbox{\boldmath $v$})=\frac{1}{T}(2\pi T)^{-\frac{d-1}{2}}|v_z|
               \exp\left(-\frac{\mbox{\boldmath $v$}^2}{2T}\right),
\label{eq:distribution}
\end{eqnarray}
being independent of incoming velocity.
Here $d$ denotes the dimension of the system and the Boltzmann constant is chosen to be 1.
In the steady state density of the particles has a finite gradient; low at the hot end and high at the cold end.
The density all over the system is chosen
so that it will be high to model the dense fluid and lower than solidified density at the densest position;
that is, 0.52 for the two-dimensional systems and 0.36 for the three-dimensional systems relative to the close-packed density.

Under these conditions, the equipartition of energy and the
absence of the hydrodynamic flow are simulationally confirmed.
We define the local temperature as
\begin{equation}
T(z)=\frac{1}{d}
\Bigl< \sum_{z-\epsilon < r^z_i < z+\epsilon} \mbox{\boldmath $v$}_i^2 \Bigr>
\Big/
\Bigl< \sum_{z-\epsilon < r^z_i < z+\epsilon} 1 \Bigr>,
\end{equation}
in the same way as the lattice case(Eq.(\ref{tempdiff})).
Both in the two and three dimensional systems,
temperature profiles have linear forms except for the area nearby boundary.
For this reason,
we use the temperature gradient obtained by the fitting in the bulk 
for the $\mbox{grad} (T)$ in Eq.(\ref{fourier})
to calculate the thermal conductivity of the system.
The global heat flux is measured as the energy received by the cold bath per unit time in the steady state.

\begin{figure}[hbtp]
\begin{center}
\epsfig{file=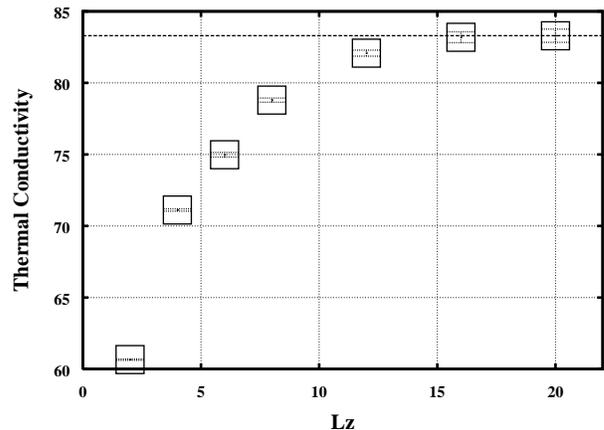,width=8.5cm}
\caption{Size dependence of thermal conductivity in three-dimensional hard core fluid
system. $\kappa(L_z)$ saturates a constant value.
Error bars show $2\sigma$ error regions.}
\label{fig:k(L)}
\end{center}
\end{figure}

Size-dependence of the thermal conductivity $\kappa$ is again investigated.
In the two-dimensional system, $\kappa(L_z)$ diverges as $\sim L_z^\alpha$ where $\alpha$ is estimated less than $0.2$.
This result is consistent with the logarithmic divergence in lattice system\cite{LIPPI}.
On the other hand, as shown in FIG.\ref{fig:k(L)}, $\kappa(L_z)$ saturates to a constant value in
the three-dimensional system.
The autocorrelation function of heat flux
in the two and three-dimensional isolated systems is also observed.
In this observation we chose the global density to the same value as the one of nonequilibrium steady
state and apply the periodic boundary condition to all direction.
The function shows power-law decay as $\sim t^{-1}$ and $\sim t^{-1.5}$ asymptotically in the two and the three-dimensional system,
respectively.

In summary,
using non-linear lattice and hard-core fluid
we have confirmed normal Fourier-type heat conduction in three-dimension.
The 1D-lattice shows the power-law divergence of the thermal conductivity as same as other one-dimensional chains.
In the two-dimensional fluid system, thermal conductivity slowly diverges in thermodynamic limit. 
In the three-dimensional systems, the thermal conductivity converges in both lattice and fluid.
We have also investigated the autocorrelation function of the heat flux 
and confirmed the long time tails whose exponent is $-d/2$.
These results mean that
the divergence behavior of thermal conductivity can be understood by the analogy with hydrodynamical system\cite{LEPRIB,LIPPI}.
It is remarkable that the quasi-one-dimensional total momentum conserving system shows the normal Fourier-type heat conduction.

The authors thank the Supercomputer Center, Institute for Solid State Physics, University of Tokyo
for the use of the FUJITSU VPP 500. 
This work is partly supported by Grants-in-Aid from the Ministry of
Education, Science, Sports and Culture (No. 11740222).


\begin{references}
\bibitem{RIEDER} Z. Rieder, J. L. Lebowitz, and E. Lieb, J. Math. Phys. {\bf 8}. 1073 (1967).
\bibitem{OCONNOR} A. J. O'Connor and J. L. Lebowitz, J. Math. Phys. {\bf 15}. 692 (1974).
\bibitem{DING-A-LING} G. Casati, J. Ford, F. Vivaldi, and W. M. Visscher, Phys. Rev. Lett. {\bf 52}, 1861 (1984). 
\bibitem{DINGDONG} G. Casati, J. Ford, F. Vivaldi, and W. M. Visscher, Phys. Rev. Lett. {\bf 52}, 1861 (1984). 
\bibitem{FK} B. Hu, B. Li and H. Zhao, Phys. Rev. Lett. {\bf 57}, 2992 (1998)
\bibitem{Lorentzgas1} J. L. Lebowitz and H. Spohn, J. Stat. Phys. {\bf 19}, 633 (1978).
\bibitem{Lorentzgas2} D. Alonso, R. Artuso, G. Casati, and I. Guarneri, Phys. Rev. Lett. {\bf 82},1859 (1999).
\bibitem{LEPRIA} S. Lepri, R. Livi, and A. Politi, Phys. Rev. Lett. {\bf 78}. 1896 (1997). 
\bibitem{Kubo1} R. Kubo, J. Phys. Soc. Jpn. {\bf 12}, No. 6, 570 (1957).
\bibitem{Kubo2} R. Kubo, M. Yokota, and S. Nakajima, J. Phys. Soc. Jpn. {\bf 12}, No. 11, 1203 (1957).
\bibitem{Alder1} B. J. Alder, D. M. Gass and T. E. Wainwright, J. Chem, Phys. {\bf 53}, 3813 (1970). 
\bibitem{Alder2} T. E. Wainwright, B. J. Alder, and D. M. Gass, Phys. Rev. A {\bf 4}, No. 1, 233 (1971). 
\bibitem{DorfmanCohen1} J. R. Dorfman and E. G. D. Cohen, Phys. Rev. A {\bf 6}, No. 2, 776 (1972).
\bibitem{DorfmanCohen2} J. R. Dorfman and E. G. D. Cohen, Phys. Rev. A {\bf 12}, No. 1, 292 (1975).
\bibitem{Ernst} M. H. Ernst, Physica D {\bf 47}, 198 (1991).
\bibitem{LEPRIB} S. Lepri, R. Livi, and A. Politi, cond-mat/9806133. 
\bibitem{HATANO} T. Hatano, Phys. Rev. E {\bf 59}. 1063 (1999).
\bibitem{LIPPI} A. Lippi and R. Livi, chao-dyn/9910034
\bibitem{NOSE} S. Nos\'{e}, J. Chem. Phys. {\bf 81}. 511 (1984).
\bibitem{HOOVER} W. G. Hoover, Phys. Rev. A {\bf 31}. 1695 (1985).
\bibitem{Ford} J. Ford, Phys. Rep. {\bf 213}. 271 (1992).
\end{references}
\end{document}